\documentclass[sigconf]{acmart}
\usepackage{enumitem}
\usepackage{multirow}
\usepackage{multicol}
\usepackage{booktabs}
\usepackage{url}
\usepackage{hyperref}
\usepackage{mathrsfs}
\usepackage{caption}
\usepackage{subfigure}
\usepackage{amsmath}
\usepackage{balance}

\newtheorem{definition}{Definition}

\copyrightyear{2023}
\acmYear{2023}
\setcopyright{acmlicensed}\acmConference[KDD '23]{Proceedings of the 29th ACM SIGKDD Conference on Knowledge Discovery and Data Mining}{August 6--10, 2023}{Long Beach, CA, USA}
\acmBooktitle{Proceedings of the 29th ACM SIGKDD Conference on Knowledge Discovery and Data Mining (KDD '23), August 6--10, 2023, Long Beach, CA, USA}
\acmPrice{15.00}
\acmDOI{10.1145/3580305.3599785}
\acmISBN{979-8-4007-0103-0/23/08}

\begin{document}

\title{ReLoop2: Building Self-Adaptive Recommendation Models \\via Responsive Error Compensation Loop}


\author{Jieming Zhu}
\authornote{Both authors contributed equally. Jieming Zhu is the corresponding author.}
\affiliation{
  \institution{Huawei Noah's Ark Lab}
  \city{Shenzhen}
  \country{China}}
\email{jiemingzhu@ieee.org}

\author{Guohao Cai}
\authornotemark[1]
\affiliation{
  \institution{Huawei Noah's Ark Lab}
  \city{Shenzhen}
  \country{China}}
\email{caiguohao1@huawei.com}

\author{Junjie Huang}
\authornote{Work done during internship at Huawei Noah's Ark Lab.}
\affiliation{
  \institution{Shanghai Jiao Tong University}
  \city{Shanghai}
  \country{China}}
\email{legend0018@sjtu.edu.cn}

\author{Zhenhua Dong}
\affiliation{
  \institution{Huawei Noah's Ark Lab}
  \city{Shenzhen}
  \country{China}}
\email{dongzhenhua@huawei.com}

\author{Ruiming Tang}
\affiliation{
  \institution{Huawei Noah's Ark Lab}
  \city{Shenzhen}
  \country{China}}
\email{tangruiming@huawei.com}

\author{Weinan Zhang}
\affiliation{
  \institution{Shanghai Jiao Tong University}
  \city{Shanghai}
  \country{China}}
\email{wnzhang@sjtu.edu.cn}
  

\renewcommand{\shortauthors}{Jieming Zhu et al.}

\begin{abstract}

Industrial recommender systems face the challenge of operating in non-stationary environments, where data distribution shifts arise from evolving user behaviors over time. To tackle this challenge, a common approach is to periodically re-train or incrementally update deployed deep models with newly observed data, resulting in a continual learning process. However, the conventional learning paradigm of neural networks relies on iterative gradient-based updates with a small learning rate, making it slow for large recommendation models to adapt. In this paper, we introduce ReLoop2, a self-correcting learning loop that facilitates fast model adaptation in online recommender systems through responsive error compensation. Inspired by the slow-fast complementary learning system observed in human brains, we propose an error memory module that directly stores error samples from incoming data streams. These stored samples are subsequently leveraged to compensate for model prediction errors during testing, particularly under distribution shifts. The error memory module is designed with fast access capabilities and undergoes continual refreshing with newly observed data samples during the model serving phase to support fast model adaptation. We evaluate the effectiveness of ReLoop2 on three open benchmark datasets as well as a real-world production dataset. The results demonstrate the potential of ReLoop2 in enhancing the responsiveness and adaptiveness of recommender systems operating in non-stationary environments. 



\end{abstract}



\maketitle

\section{Introduction}
Nowadays, personalized recommendation has emerged as a prominent channel across a range of online applications, including e-commerce, news feeds, and music apps. It enables the delivery of tailored items to users based on their individual interests. The provision of high-quality recommendations not only enhances user engagement but also fuels revenue growth for platforms.
To achieve accurate recommendations, deep learning-based models have gained widespread adoption in industry owing to their flexibility and ability to capture intricate user-item relationships. However, industrial recommender systems often operates in non-stationary environments, where data distribution shifts~\cite{distshift_survey} occur as a result of evolving user behaviors over time. This can lead to the deterioration of well-trained recommendation models during online serving, and thus poses a challenge for models to quickly adapt under distribution shifts.


To address this challenge, previous research efforts have been made from two aspects: {behavior sequence modeling} and {incremental model training}. The first line of research aims to capture dynamic patterns at the feature level by modeling sequential user behavior sequences. Notable progress has been made in this area, particularly in click-through rate (CTR) prediction tasks~\cite{FuxiCTR}, with models incorporating attention, GRU, and transformer architectures, such as DIN~\cite{DIN}, DIEN~\cite{DIEN}, and BST~\cite{BST}. These models formulate CTR prediction as a few-shot learning task~\cite{Fewshot_survey}, where given k historical behaviors from a user (i.e., k-shot samples), the goal is to predict whether the user will click on the next item. Few-shot learning enables rapid adaptation to new users with only a few observed samples~\cite{FewShotRec}. However, these studies do not explicitly address the test-time distribution shift problem. In parallel, another line of research focuses on incremental model training. While regular re-training of models (e.g., daily) is straightforward, it becomes time-consuming due to the large volume of training data in practice (e.g., up to billions of samples in Google Play's app recommendation~\cite{WideDeep}). As a result, incremental model training~\cite{IncrementalCTR,IncCTR,IncrementalRS} has gained popularity in industrial recommender systems. This approach retains previous model parameters for initialization~\cite{IncrementalCTR} or knowledge transfer~\cite{IncrementalRS} while continually updating the model using newly observed data samples, often at a minute-level granularity. Incremental training brings training efficiency and enhances model freshness. However, learning the parameters of neural networks relies on iterative gradient-based updates to gradually incorporate supervision information into model weights with a small learning rate. This makes it challenging for large parametric recommendation models to achieve fast adaptation to distribution shifts. This challenge, often referred to as the stability-plasticity dilemma~\cite{Dilemma} in incremental learning, stems from the need to balance the stability of existing knowledge with the plasticity required to incorporate new information efficiently.


In comparison, humans possess an extraordinary capability for fast incremental learning in dynamic environments~\cite{CLS3}. 
This remarkable learning ability can be attributed to the presence of two complementary learning systems (CLS) in the human brain: the \textit{hippocampus} and the \textit{neocortex}~\cite{CLS1,CLS2}. The hippocampus is responsible for rapid learning of recent specific experiences and exhibits short-term adaptation. On the other hand, the neocortex functions as a slow learning system, gradually acquiring structured knowledge about the environment over time. The combination of these slow and fast CLS learning mechanisms empowers humans to learn quickly and remember information in the long term. This inspires us to design an adaptive recommendation framework that incorporates both fast and slow learning modules to build self-adaptive recommendation models and make accurate recommendations in dynamic environments.

Our approach consists of two key components: a base model serving as a slow learning module that updates through gradient back-propagation, and a fast learning module equipped with a non-parametric error memory. Unlike traditional gradient-based training, the fast learning module is training-free and thus enables rapid adaptation to new data distributions. To be specific, consider the learning process of students, who often organize and review their past incorrect questions to reflect on their errors and improve their performance in subsequent exams. Inspired by this, we propose to store recent error samples from the incoming data stream in the error memory. These error samples reflect situations where the model's performance degrades, particularly during distribution shifts. In response, we estimate the errors between the base model's predictions and the ground-truth labels using the error memory. These error estimations are then utilized to compensate for the model's degradation caused by distribution shifts. During the model serving phase, newly observed data is continuously written to the error memory, creating a self-correcting learning loop that facilitates fast model adaptation. We refer to this approach as ReLoop2, which extends the original ReLoop learning framework~\cite{ReLoop} to test-time adaptation.

In building a self-correcting learning loop for online recommendation, we encounter two primary challenges. The first challenge pertains to estimating the errors for output compensation without relying on back-propagation. To address this, we propose a non-parametric method that leverages the error memory to retrieve similar samples, enabling us to approximate the errors effectively. The second challenge lies in the time- and space-efficient design of the error memory. Given the substantial volume and high velocity of data streams, we introduce a locality-sensitive hashing (LSH)-based sketching approach. This approach ensures efficient O(1)-time memory reading and writing operations while maintaining a constant memory footprint.


The ReLoop2 framework establishes a continual model adaptation process by continuously refreshing the error memory with newly observed error samples after model deployment. It has the potential to significantly enhance model performance when faced with data distribution shifts. Importantly, the framework is orthogonal to existing incremental learning techniques and compatible with diverse models used in recommendation systems. We empirically validate the effectiveness of ReLoop2 on three open benchmark datasets and a proprietary production dataset, showcasing substantial performance improvements over existing models and incremental training techniques. We hope that our work could inspire further research attention to address the challenge of train-test distribution shift in online recommender systems. In summary, this paper makes three main contributions:
\begin{itemize}
    \item We identify the challenging problem of fast model adaptation for online recommendation, and propose a slow-fast learning paradigm inspired by the complementary learning systems observed in human brains.
    \item We introduce a time- and space-efficient non-parametric error memory and leverage it to build a responsive error compensation loop for fast model adaptation.
    \item We conduct extensive experimental evaluations on both open benchmark datasets and real-world production datasets to demonstrate the effectiveness of our ReLoop2 framework.
\end{itemize}

\section{Background}
In this section, we provide an overview of the CTR prediction task. We then describe our motivation for fast model adaptation in real-world scenarios. Additionally, we review the locality-sensitive hashing (LSH) technique used in our work.

\subsection{CTR Prediction}\label{sec:ctr}
Typically, input samples consist of two main types of features: categorical features, and numerical features. In our approach, we utilize embedding techniques to map these features into a lower-dimensional embedding space. Specifically, for numerical features, we first discretize them into categorical bucket features. Then, same with categorical features, we apply one-hot/multi-hot encoding and embedding table lookups to obtain embedding vectors. Let $x = \{x_1,x_2,...,x_n\}$ denotes a data instance with $n$ feature fields. Then we could get its feature embeddings as $e = \{e_1,e_2,...,e_n\}$, which serve as the input to a deep neural network.


Given the set of feature embeddings, various CTR prediction models have been proposed to model feature interactions (e.g., DeepFM~\cite{DeepFM}, DCN~\cite{DCN}, DCN-V2~\cite{DCN_V2}) and sequential user interests (e.g., DIN~\cite{DIN}, DIEN~\cite{DIEN}, and BST~\cite{BST}). At last, the CTR model outputs the predicted click probability $\hat{y} \in [0,1]$ using the sigmoid activation $\sigma(\cdot)$ on the logit value. Formally, we denote a CTR prediction model as follows:
\begin{equation}
\hat{y} = \sigma \big(\phi(e) \big)
\end{equation}
where $\phi(\cdot)$ is a multi-layer deep neural network. For example, $\phi_{DeepFM}$, $\phi_{DCN}$, and $\phi_{DIN}$ are commonly used model architectures~\cite{DeepFM,DCN,DIN}.


We denote $y \in \{0, 1\} $ as a true label to indicate whether a user has clicked a recommended item. The binary cross-entropy loss function is usually used for CTR prediction:
\begin{equation}
    \textit{L} =-\frac{1}{N} \sum \big(ylog\hat{y} + (1-y)log(1-\hat{y}) \big)
\end{equation}
where $N$ is the number of training instances. Readers may refer to our BARS benchmark~\cite{FuxiCTR} for more training details.


\subsection{Fast Model Adaptation}


\begin{figure}[!b]
    \centering
     \includegraphics[scale=0.25]{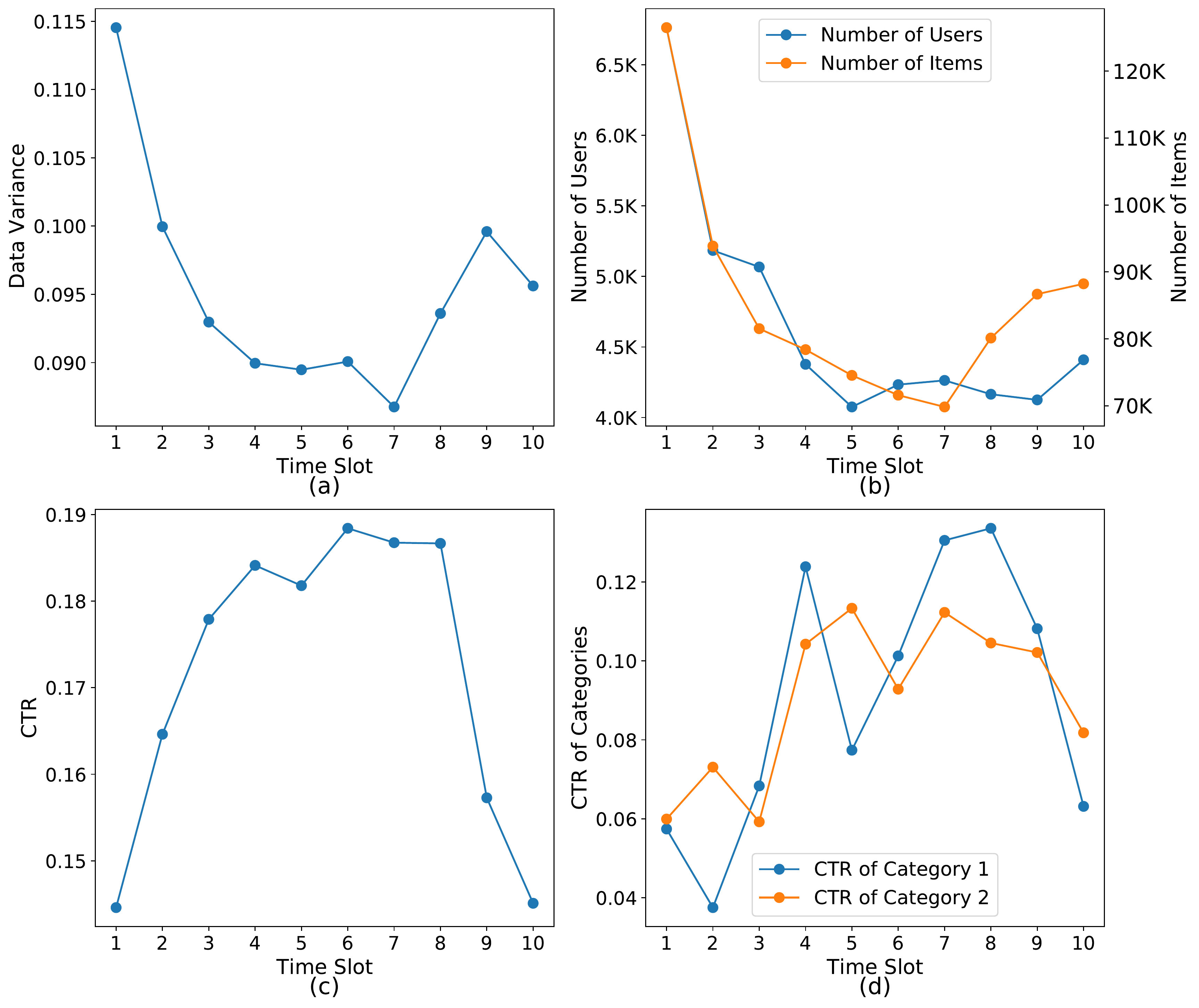}
    \caption{Observations of data distribution shifts on the MicroVideo dataset. (a) Variance of data samples over each time slot. (b) Number of users and items involved over each time slot. (c) The averaged CTR over each time slot. (d) The averaged CTR of  two unique categories over each time slot.}
    \label{fig:motivation}
\end{figure}

In this section, we analyze the motivation behind the need for fast model adaptation to address the problem of distribution shift. In Figure~\ref{fig:motivation}, we present our observations regarding the dynamic data distribution from various perspectives, including data variance, feature dynamics, overall CTR, and category-specific CTR over time. To illustrate this, we split the test dataset of MicroVideo (detailed in Section~\ref{exp}) into ten chronological time slots, simulating an online advertising scenario. Specifically, Figure~\ref{fig:motivation}(a) depicts the data variance (from embedding $e$) for each time slot, revealing the spread of data instances relative to their average. A higher value indicates a greater deviation from the average. Figure~\ref{fig:motivation}(b) demonstrates the changes in the number of users and items over time. Both (a) and (b) highlight the covariate shift in feature $x$. Furthermore, Figure~\ref{fig:motivation}(c) showcases the dynamic nature of CTR over time, while (d) exhibits the dynamic CTR based on different categories. These figures reveal the label shift in $y$ and the concept drift between $x$ and $y$, respectively. Collectively, these visualizations demonstrate a significant level of data change occurring over time. As a result, there is a pressing demand for fast model adaptation to swiftly adjust to the dynamic patterns present in the data.



\subsection{Locality Sensitive Hashing}\label{sec:lsh}
This section provides a brief review of the classical Locality Sensitive Hashing (LSH) algorithm~\cite{LSH1,LSH2}, which is a widely adopted sublinear-time algorithm for approximate nearest neighbor search. In LSH, a hash function $h(x)\mapsto \mathbb{Z}$ a mapping that assigns an input $x$ to an integer in the range $[0, R-1]$. LSH encompasses a family of such hash functions with a key property: similar points have a high probability of being mapped to the same hash value~\cite{LSH2}. More formally, a LSH family is defined as follows~\cite{LSH2}.

\begin{definition} \label{def:lsh}{\bf LSH Family}.~~A family $\mathcal{H}$ is called $(S_0,cS_0,p_1,p_2)$-sensitive with respect to a similarity measure $sim(\cdot,\cdot)$ if for any two points $x,y \in \mathbb{R}^d$ and $h$ chosen uniformly from $\mathcal{H}$ the following properties hold:
\begin{itemize}
\item if $sim(x,y)\ge S_0$ then $p(x,y) \ge p_1$
\item if $sim(x,y)\le cS_0$ then $p(x,y) \le p_2$
\end{itemize}
 \end{definition}

Typically, $p_1 > p_2$ and $c < 1$ is required for approximate nearest neighbor search. We use the notation $p(x,y)$ to denote the collision probability ${Pr}\big(h(x)=h(y)\big)$ between $x$ and $y$, where their hash values are equal. One sufficient condition for being a LSH family is that the collision probability $p(x,y)$ is a monotonically increasing function of similarity between the two data points, i.e.,
\begin{equation}
p(x,y) = f\big(sim(x, y)\big)
\end{equation}
where $f(\cdot)$ is required to be a monotonically increasing function. In other words, similar data points are more likely to collide with each other under LSH mapping.

Among the widely known LSH families, SimHash~\cite{SimHash2} is a popular LSH that applies the technique of Signed Random Projections (SRP)~\cite{SimHash1,SimHash2,SimHash3} for the cosine similarity measure. Given a vector $x$, SRP utilizes a random $w$ vector with each component generated from i.i.d. normal, i.e., $w_i \sim N(0, 1)$, and only stores the sign of the projection. Hence, SimHash is given by 
\begin{equation}\label{equ:sign}
h_w(x) = sign(w^Tx)~.
\end{equation}
Particularly, we could take $[h_w(x)]_+=max\big(0, h_w(x)\big)$ using the ReLU function to re-map it to $\{0, 1\}$. It has been shown in the seminal work~\cite{SimHash3} that the collision probability under SRP satisfies the following equation:
\begin{equation}\label{p_sign}
p(x,y)= 1 - \frac{1}{\pi}cos^{-1}\big(\frac{x^Ty}{\left \|x\right \|_2\left \|y\right \|_2}\big)
\end{equation}
where $p(x,y)$ is monotonic to the cosine similarity $\frac{x^Ty}{\left \|x\right \|_2\left \|y\right \|_2}$.

It is important to note that each hash function $h_w(x)$ generates a single bit using SRP, resulting in two possible hash values $\{0, 1\}$. By independently sampling $L$ hash functions with different $w$ vectors, we can generate new hash values in the range $[0, 2^L-1]$ by combining the outcomes of the $L$ independent SRP bits. The collision probability is equal to $p(x, y)^L$, the power of $L$ of Equation~\ref{p_sign}.

\begin{figure*}[!thbp]
	\centering
	\includegraphics[width=0.9\textwidth]{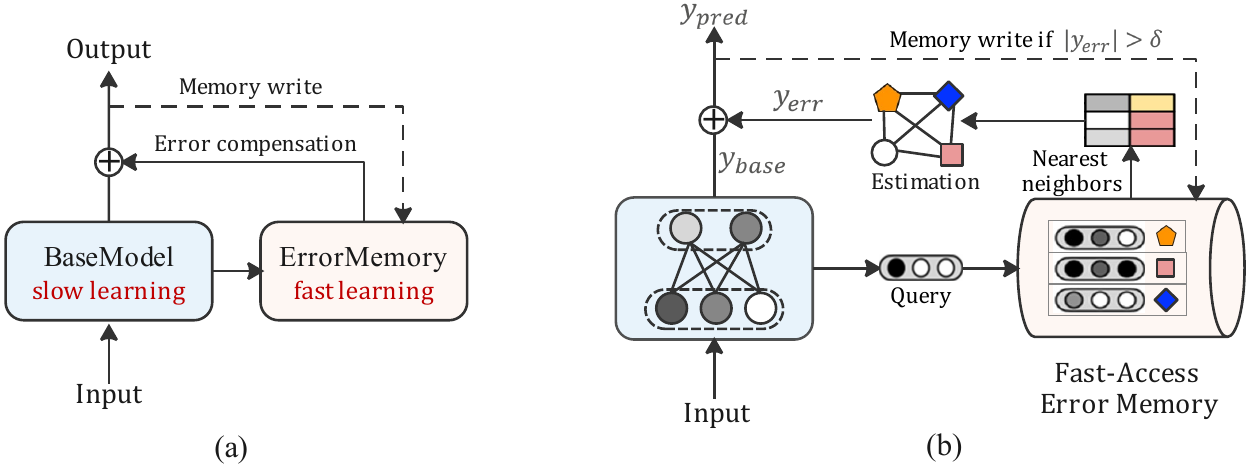}
	\caption{
	(a) An overview of our slow-fast learning framework, which comprises  a slow-learning base model and a non-parametric error memory module for fast adaptation; (b) The ReLoop2 diagram that builds a self-correcting learning loop with error compensation. 
	}
	\label{fig:framework}
\end{figure*}

\section{Approach}
In this section, we present our ReLoop2 apporach that enables fast model adaptation through a self-correcting learning loop with responsive error compensation.

\subsection{Overview}
Deep learning-based recommendation models, such as CTR prediction models discussed in Section \ref{sec:ctr}, are typically optimized using back-propagation algorithms within the empirical risk minimization (ERM) framework. These models assume a stationary data distribution (i.e., the training and testing data are drawn from the same distribution) and require a small learning rate to gradually incorporate information into model weights. However, in real-world online recommendation scenarios, the rapid emergence of new users and items, along with potential changes in user behavior over time, result in the distribution shift challenge. Consequently, a well-trained model may gradually degrade after deployment. To address this challenge, we propose the ReLoop2 framework for fast model adaptation, as depicted in Figure~\ref{fig:framework}.

Our framework employs a slow-fast learning paradigm, where the base model undergoes slow gradient updates, while an episodic memory module, free from back-propagation, is introduced to facilitate rapid acquisition of new knowledge. The base model is a standard parametric neural network that learns common knowledge through gradual gradient updates. In contrast, the memory module is a non-parametric component that stores recently observed samples and enables fast learning and adaptation from these samples. This slow-fast learning paradigm aligns with the theory of complementary learning systems (CLS) in human brains~\cite{CLS1,CLS2}. 

Specifically, we refer to the memory module as the error memory, which stores the recent error samples produced by the base model on the incoming data stream. These error samples provide insights into cases where the model makes incorrect predictions, particularly in the presence of distribution shifts. By directly capturing and remembering these error samples, we can estimate errors in a non-parametric manner and subsequently correct the model output through error compensation. This establishes a continual fast adaptation process for the model within the evolving dynamics of the non-stationary data distribution. New error samples observed during model deployment are continuously written back to the error memory, enabling the tracking of changing dynamics in the online data.

\subsection{Error Compensation Loop}
Figure~\ref{fig:framework}(b) depicts our error compensation loop for fast model adaptation, comprising three key components: the base model $\phi$, the fast-access error memory $M$, and the error estimation module $\mathcal{E}$. Our learning framework is generic and compatible with various base models used for CTR prediction. We formulate the base model as follows:
\begin{equation}
y_{base} = \phi(e)
\end{equation}
where $e$ represents the feature embeddings of a data instance. $y_{base} \in [0, 1]$ denotes the predicted click probability from the base model. We provide a brief overview of feature embedding and CTR modeling approaches in Section~\ref{sec:ctr}. It is worth noting that the model function $\phi$ can be implemented using any existing CTR prediction model, such as DeepFM~\cite{DeepFM}, DCN~\cite{DCN}, and DIN~\cite{DIN}. 
The base model approximates the ground truth label $y$ by minimizing the error between $y$ and $y_{{base}}$ within an empirical risk minimization framework: 
\begin{equation}\label{equ:error}
\min \enspace \epsilon^2 \text{~,\quad where} \enspace \epsilon = y - y_{base}
\end{equation}
Ideally, under the assumption of independent and identically distributed (i.i.d.) data, the error $\epsilon$ should be a small random variable close to zero after model training. However, the base model degrades when confronted with distribution shifts, resulting in an enlarged error $\epsilon$ during model serving.

To address this issue, we propose a proactive approach to estimate the model prediction error and correct the model output accordingly. However, directly obtaining $\epsilon$ from Equation~\ref{equ:error} is not feasible due to the unknown label $y$ during prediction. Therefore, we propose to estimate it using recently observed samples stored in the error memory module $M$. Formally, we perform error estimation with the following formula:
\begin{equation}\label{equ:estimation}
y_{err} = \mathcal{E}(h_q, M)
\end{equation}
where $h_q$ denotes the hidden representation of input sample $q$, which can be chosen from any hidden layer (e.g., the last hidden layer) of the base model $\phi$. With the estimated error $y_{err}$, we can make compensation for the model output to correct its prediction:
\begin{equation}\label{equ:compensation}
y_{pred} = y_{base} + \lambda \cdot y_{err}
\end{equation}
where $y_{pred}$ denotes the final output with model adaptation. The compensation weight $\lambda$ adjusts the proportion of error compensation. It is important to note that the value of $y_{pred}$ may exceed the range $[0, 1]$ after error compensation. In such cases, we clamp the value within the range.

In the following sections, we will describe our designs for the error estimation module $\mathcal{E}$ and the error memory module $M$.

\subsubsection{Error Estimation Module}
Given the error memory that retains recently observed data samples, our goal is to estimate the prediction error based on similar samples to a new input $q$. Formally, we aim to retrieve a set of top-k similar samples from the memory, as described below:
\begin{equation}\label{equ:retrival}
\mathcal{K}=\big\{(s_i, ~y_i, ~y_{base\_i}) ~|~ i \in M\big\}
\end{equation}
where $s_i=sim(h_q, h_i)$ denotes the similarity between the hidden vectors of the query sample $x$ and memory instance $i$. Additionally, $y_i$ and $y_{base\_i}$ represent the ground-truth label and the prediction value of the base model, respectively. The derivation of similar samples $\mathcal{K}$ is provided in Section~\ref{sec:memory}.

Inspired by the attention mechanism employed in content-addressing memory networks~\cite{memorynet}, we can estimate the attention-weighted ground truth $\bar{y}$ and prediction value $\bar{y}_{base}$ as follows.
\begin{equation}
 \bar{y}  = \sum\nolimits_{i \in \mathcal{K}} a_i \cdot y_i~, \enspace\enspace \bar{y}_{base} = \sum\nolimits_{i \in \mathcal{K}} a_i \cdot y_{base\_i}
\end{equation}
The attention weight $a_i$ is computed using the following equation:
\begin{equation}
a_i = \frac{exp \big(s_{i}/\tau \big)}{\sum_{i \in \mathcal{K}} exp(s_{i}/\tau)}
\end{equation} 
Here, $\tau$ is a temperature parameter that adjusts the smoothness of the softmax. The value of $\tau$ can be learned jointly with the base model or manually tuned as a hyper-parameter (e.g., $0.1$).

Next, we estimate the prediction error as a weighted combination of two error measures:
\begin{eqnarray}\label{equ:err}
y_{err} &=& \gamma \cdot (\bar{y} - y_{base}) + (1-\gamma) \cdot (\bar{y}_{base} - y_{base}) \nonumber \\
&=& \gamma \cdot\bar{y} + (1-\gamma)\cdot  \bar{y}_{base} -  y_{base} 
\end{eqnarray}
where $\gamma$ is a weight that balances the two error measures. Notably, when $\gamma=0$, the error is estimated from the model predictions on similar samples. When $\gamma=1$, the error is computed from the ground truth labels of similar samples. In the latter case, we substitute $y_{err}$ into Equation~\ref{equ:compensation} and can obtain the corrected model prediction with error compensation as follows:
\begin{eqnarray}
y_{pred} &=& y_{base} + \lambda\cdot(\bar{y} - y_{base}) \nonumber \\
&=& (1 - \lambda)\cdot y_{base} + \lambda\cdot \bar{y}
\end{eqnarray}
This can be seen as a weighted ensemble of the base model's output $y_{base}$ and the estimation $\bar{y}$ from k-nearest neighbors (KNN). For simplicity, we use $\gamma=1$ in our experiments.


\subsubsection{Fast-Access Error Memory}\label{sec:memory}
In this section, we focus on the key component of our framework, the error memory $M$. In online recommendation systems, data is acquired sequentially over time, and the model generates click predictions based on the received samples. The true labels are received when users interact with the recommended items. During this process, we can easily obtain the hidden representation $h_i$ from a specific hidden layer of the base model (same with $h_q$), the base model output $y_{base\_i}$, and the true label $y_i$. To enable fast adaptation to distribution shifts, we utilize the memory to store these recently observed samples. Ideally, our memory consists of a set of key-value pairs formulated as follows:
\begin{equation}\label{equ:mem}
M = \big\{(h_i, ~y_i, ~y_{base\_i})\big\}
\end{equation}
where $h_i \in \mathbb{R}^d$ represents the d-dimensional key vector of memory slot $i$, while $y_i$ and $y_{base\_i}$ serve as the memory values. We define a memory reader function $R$ that retrieves a set of similar samples $\mathcal{K}$ from the memory given $h_q$ as a query:
\begin{equation}\label{equ:reader}
\mathcal{K} = R(h_q, ~M)
\end{equation}

However, designing the memory poses two main challenges in real-world recommender systems due to the large volume of click data:
\begin{itemize}
    \item \textbf{Fast Access}. For real-time online CTR prediction, model inference must meet stringent latency requirements. Therefore, it is crucial to enable fast access to the error memory. However, retaining a large number of recently observed data samples for adaptation makes the memory size too large to utilize traditional attention-based content addressing mechanisms in memory networks~\cite{memorynet}, which needs to read all memory slots for each query.
    \item \textbf{Memory Size}. The memory module requires a substantial memory size to store an adequate number of data samples. In our production system, the number of samples can easily reach millions within a 10-minute timeframe. Storing such a large number of data samples consumes significant computing resources (e.g., RAM) for model serving. Therefore, minimizing memory resource consumption for the error memory is highly desirable.
\end{itemize}

To address these challenges, we explore two potential solutions: approximate nearest neighbor (ANN) search and LSH-based sketching. ANN search techniques are widely used in industry to efficiently retrieve top-k nearest vectors in sub-linear time. These techniques have been successful in various retrieval-augmented machine learning tasks (e.g., language modeling~\cite{knn_LM}, machine translation~\cite{knn_translation}). Additionally, they are supported by mature tools and libraries, including Faiss~\cite{faiss}, ScaNN~\cite{scann} and Milvus~\cite{milvus}. However, constructing popular ANN indices like HNSW~\cite{hnsw} and IVFPQ in Faiss~\cite{faiss} involves time-consuming steps (e.g., k-means) and requires substantial memory (gigabytes of RAM). To reduce memory consumption, we propose filtering the data samples based on the model's errors. Specifically, we only store samples with relatively large errors (greater than a threshold $\sigma$) since they indicate significant degradation in the base model. Despite applying error filtering and random down-sampling, storing these raw data samples still imposes a considerable burden on memory. Therefore, for online recommendation tasks with limited computing resources, ANN search may not be the optimal choice.

Ideally, we aim to design a lightweight and fast-access memory that avoids directly storing massive data points in RAM, eliminates the need for iterative and non-streaming processes like k-means, and avoids constructing complex index structures such as graphs, which are either time-consuming or difficult to parallelize. To achieve these objectives, we propose an alternative design of the error memory by employing the LSH-based data sketching algorithm on the streaming data~\cite{RACE}. LSH, as introduced in Section~\ref{sec:lsh}, enables efficient bucketing of each data point in constant time using fixed hash functions. Data sketching supports the construction of a compact sketch that summarizes the streaming data. The sketching algorithm compresses a set of high-dimensional vectors into a small array of integer counters, which is sufficient for estimating the similarity $s_i$ of similar samples in Equation~\ref{equ:reader}.

Formally, we define the memory as a sketch consisting of $K$ repeated arrays, denoted as $M_k$ for $k \in [0, K-1]$. Each array $M_k$ is indexed by $L$ independent hash functions $H_L(x)=\{h_w(x) ~|~ w\}$, where $h_w(x)\mapsto [0, 1]$ is a singed random projection described in Equation~\ref{equ:sign}. Consequently, an input $x$ can be hashed into an index in $2^L$ buckets: $H_L(x) \mapsto [0, 2^L-1]$. For example, setting $L=16$ can result in approximately 65,536 buckets. While the sketch is originally designed for kernel density estimation with integer counters~\cite{RACE}, in our design, we store a tuple of summation values at bucket $b$ in the array $M_k$, denoted as $M_k[b]=(sum\_x[b], sum\_y[b], sum\_y_{base}[b])$. To ensure more stable estimations, the same process is repeated $K$ times using $K$ different sets of hash functions $\{H_L^k(x) ~|~ k \in [0, K-1]\}$.  In summary, the memory can be viewed as a concatenated array of size $2^L \times K \times 3$. More specifically, we formulate the memory writing and reading processes as follows:

\textbf{Memory Writing}. For each observed sample $i$ from the data stream, we obtain  $(h_i, y_i, y_{base\_i})$. Instead of directly storing the raw samples in the memory following Equation~\ref{equ:mem}, we apply each set of hash functions $H_L^k$ to map the key vector $h_i$ to the corresponding bucket $b$ and update the sketch array $M_k[b]$ as follows.
\begin{equation}
\begin{aligned}
sum\_x[b] ~&+=~ 1 \\
sum\_y[b] ~&+=~ y_i \\
sum\_y_{base}[b] ~&+=~ y_{base\_i}
\end{aligned}
\end{equation}\label{equ:write}
Note that the values in $M_k[b]$ are initially set to zero and can be reset regularly or when the base model has been updated to refresh the memory. The updates on all $K$ memory arrays can be performed in parallel.

\textbf{Memory Reading}. Given a query sample vector $h_q$, we can apply the same set of hash functions to map the query to bucket $b$. We then obtain the summation values from each sketch array $M_k[b]$ and compute the readout values via averaging them over all buckets as follows:
\begin{equation}\label{equ:read}
\begin{aligned}
s_i ~&=~ sum\_x[b] ~\big/~ \sum\nolimits_{b} sum\_x[b] \\
y_i ~&=~ sum\_y[b] ~\big/~ \sum\nolimits_{b} sum\_x[b]   \\
y_{base\_i} ~&=~ sum\_y_{base}[b] ~\big/~ \sum\nolimits_{b} sum\_x[b]
\end{aligned}
\end{equation}
After parallel reading from $K$ sketch arrays, we obtain the $K$ readout results of similar samples $\mathcal{K} = \big\{(s_i, y_i, y_{base\_i}) ~|~ i \in [0, K-1]\big\}$, which can then be used in Equation~\ref{equ:retrival} for error estimation.

Compared to traditional memory that stores raw samples, our LSH-based sketch memory offers several advantages.  It enables fast construction time (O(1) writing time per sample), has a low memory requirement (constant memory size of $2^L \times K \times 3$), and eliminates the need for query-time distance computations (O(1) reading time per query). It is worth noting that our sketch memory is not only practical to implement but also enjoys strong theoretical guarantees~\cite{RACE}.

In this way, the error memory module helps estimate the potential error of the base model based on observed similar data samples, contributing to an error compensation loop that continuously and adaptively corrects the model output. This results in a slow-fast joint learning framework for fast model adaptation.



\section{Experiments}\label{exp}

\begin{table*}[!t]
\renewcommand\arraystretch{1.0}
\setlength{\tabcolsep}{2.5pt}
\caption{Performance comparison against the state-of-the-art models. BASE+ReLoop2 represents the setting that ReLoop2 is integrated to the best performing base model on each dataset.}
\resizebox{\textwidth}{!}{%
\begin{tabular}{ccccccccccccc}
\toprule
\multirow{2}{*}{Model} & \multicolumn{4}{c}{AmazonElectronics} & \multicolumn{4}{c}{MicroVideo} & \multicolumn{4}{c}{KuaiVideo}  \\
\cmidrule(lr){2-5}\cmidrule(lr){6-9}\cmidrule(lr){10-13}\
 & gAUC(\%) & RelImp & AUC(\%) & RelImp & gAUC(\%) & RelImp & AUC(\%) & \multicolumn{1}{l}{RelImp} & gAUC(\%) & RelImp & AUC(\%) & RelImp  \\ \midrule
FM & 84.94 & \multicolumn{1}{c}{0\%} & 84.85 & \multicolumn{1}{c}{0\%} & 67.24 & \multicolumn{1}{c}{0\%} & 71.86 & \multicolumn{1}{c}{0\%} & 65.99 & \multicolumn{1}{c}{0\%} & 74.18 & 0\% \\
FmFM & 85.29 & 0.4\% & 85.47 & 0.7\% & 67.37 & 0.2\% & 72.15 & 0.4\% & 65.52 & -0.7\% & 73.89 & -0.4\% \\
DeepFM & 87.89 & 3.5\% & 88.16 & 3.9\% & 68.52 & 1.9\% & 73.37 & 2.1\% & 66.65 & 1.0\% & 74.52 & 0.5\% \\
DCN & 87.78 & 3.3\% & 88.01 & 3.7\% & 68.55 & 1.9\% & 73.42 & 2.2\% & 66.58 & 0.9\% & 74.61 & 0.6\% \\
xDeepFM & 87.90 & 3.5\% & 88.13 & 3.9\% & \underline{68.89} & \underline{2.5\%} & \underline{73.62} & \underline{2.4\%} & 66.96 & 1.5\% & 74.71 & 0.7\% \\
AutoInt+ & 87.87 & 3.4\% & 88.04 & 3.8\% & 68.46 & 1.8\% & 73.38 & 2.1\% & 66.67 & 1.0\% & 74.69 & 0.7\% \\
DCNv2 & 87.90 & 3.5\% & 88.12 & 3.9\% & 68.59 & 2.0\% & 73.44 & 2.2\% & 66.75 & 1.2\% & 74.70 & 0.7\% \\
AOANet & 87.91 & 3.5\% & 88.12 & 3.9\% & 68.58 & 2.0\% & 73.46 & 2.2\% & 66.79 & 1.2\% & 74.70 & 0.7\% \\
DIN & 88.35 & 4.0\% & 88.60 & 4.4\% & 68.83 & 2.4\% & 73.60 & 2.4\% & 66.96 & 1.5\% & 74.95 & 1.0\% \\
DIEN & \underline{88.56} & \underline{4.3\%}  & \underline{88.88}   & \underline{4.7\%} & 68.67 & 2.1\% & 73.21 & 1.9\% & \underline{67.11} & \underline{1.7\%} &  \underline{75.04} & \underline{1.2\%} \\
BST & 88.41 & 4.1\% & 88.64 & 4.5\% & 68.54 & 1.9\% & 73.42 & 2.2\% & 66.90 & 1.4\% & 74.84 & 0.9\% \\ \midrule
BASE+ReLoop2 & \textbf{89.33} & \textbf{5.2\%} & \textbf{89.62} & \textbf{5.6\%} & \textbf{69.53} & \textbf{3.4\%} & \textbf{74.11} & \textbf{3.1\%} & \textbf{67.18} & \textbf{1.8\%} & \textbf{75.13} & \textbf{1.3\%} \\ \bottomrule
\end{tabular}
}
\vspace{0ex}
\label{tab:main_table}
\end{table*}

In this section, we present extensive experimental results conducted on three open benchmark datasets and one real-world production dataset to validate the effectiveness of ReLoop2. Our experiments aim to answer the following three research questions:
\begin{itemize}[leftmargin=*]
    \item \textbf{RQ1}: How does the integration of ReLoop2 with state-of-the-art models contribute to the improvement of model performance?
    \item \textbf{RQ2}: How does ReLoop2 compare to incremental training in terms of performance?
    \item \textbf{RQ3}: How do different hyperparameters affect model performance?
\end{itemize}

\subsection{Experimental Setup}
\textbf{Datasets.} We conduct experiments on three open benchmark datasets, and a large-scale production dataset. 
\begin{itemize}[leftmargin=*]
\item \textbf{AmazonElectronics} is a subset of the Amazon dataset~\cite{Amazon}, a widely used benchmark dataset for recommendation. We follow the DIN work~\cite{DIN} to preprocess the dataset. Specifically, the AmazonElectronics contains 1,689,188 samples, 192,403 users, 63,001 goods and 801 categories. Features include goods\_id, category\_id, and their corresponding user-reviewed sequences: goods\_id\_list, category\_id\_list.

\item \textbf{MicroVideo} is an open dataset for short video recommendation, which has been released by~\cite{MicroVideo}. We follow the same preprocessing steps. It contains 12,737,617 interactions that 10,986 users have made on 1,704,880 micro-videos. The labels include click or non-click, while the features include user\_id, item\_id, category, and the extracted image embedding vectors of cover images of micro-videos.

\item \textbf{KuaiVideo} is another open dataset for short video recommendation. We follow the work~\cite{DBLP:conf/mm/LiLYCXN19} to obtain the preprocessed dataset. Specifically, we randomly selected 10,000 users and their 3,239,534 interacted micro-videos. It contains several interaction data between users and videos, such as user\_id, photo\_id, duration\_time, click, like, and so on. In addition, 2048-dimensional video embeddings are provided as content features.


\item \textbf{Production} is a production dataset from Huawei's news feed recommendation. It has a total of 500 million records sampled from 7 days user logs and each record has more than 100 fields of features, such as doc\_id, category\_id, short-term interest topic\_id, and anonymous data masking user\_id. We use the latest 2-hour samples as testing data, and split it into 12 consective parts in chronological order.

\end{itemize}

\begin{table}[!t]
\renewcommand\arraystretch{1.2}
\setlength{\tabcolsep}{3.4pt}
\caption{Evaluation of ReLoop2 across different base models.}
\scalebox{0.92}
{
\begin{tabular}{ccccccccc}
\toprule
\multirow{3}{*}{Model} & \multicolumn{4}{c}{AmazonElectronics} & \multicolumn{4}{c}{MicroVideo} \\ \cmidrule(lr){2-5}\cmidrule(lr){6-9}
 & \multicolumn{2}{c}{Base} & \multicolumn{2}{c}{+ReLoop2} & \multicolumn{2}{c}{Base} & \multicolumn{2}{c}{+ReLoop2} \\ \cmidrule(lr){2-5}\cmidrule(lr){6-9}
 & gAUC & AUC & gAUC & AUC & gAUC & AUC & gAUC & AUC \\ \hline
xDeepFM & 87.90 & 88.13 & \textbf{88.73} & \textbf{88.96} & 68.89 & 73.62 & \textbf{69.53} & \textbf{74.11} \\
DCNv2 & 87.90 & 88.12 & \textbf{88.81} & \textbf{88.99} & 68.59 & 73.44 & \textbf{69.71} & \textbf{74.25} \\
AOANet & 87.91 & 88.12 & \textbf{88.75} & \textbf{88.94} & 68.58 & 73.46 & \textbf{69.11} & \textbf{73.86} \\
DIN & 88.35 & 88.60 & \textbf{89.10} & \textbf{89.34} & 68.83 & 73.60 & \textbf{69.22} & \textbf{73.86} \\
DIEN & 88.56 & 88.88 & \textbf{89.33} & \textbf{89.62} & 68.67 & 73.21 & \textbf{69.15} & \textbf{73.60} \\
 \bottomrule
\end{tabular}
}
\label{tab:agnostic}
\end{table}

\textbf{Base models.} We compare our model with the following mainstream base models for CTR prediction.
\begin{itemize}[leftmargin=*]
\item Shallow models: FM~\cite{FM}, FmFM~\cite{DBLP:conf/www/SunPZF21}.
\item Feature interaction models: DeepFM ~\cite{DeepFM}, xDeepFM~\cite{xDeepFM}, DCN~\cite{DCN}, AutoInt+~\cite{autoint}, DCNv2~\cite{DCN_V2}, AOANet~\cite{autoint}.
\item Behavior sequence models: DIN~\cite{DIN}, DIEN~\cite{DIEN}, BST~\cite{BST}.
\end{itemize}

\textbf{Metrics.} We adopt the most popular ranking metrics, AUC~\cite{WideDeep} and gAUC~\cite{DIN} (i.e., user-grouped AUC), to evaluate the model performance. In addition, we report the relative improvements (RelImp) over the classic factorization machine (FM) model.

We note that the preprocessed datasets and evaluation settings for all the baseline models we studied are available on the BARS benchmark website: \textcolor{magenta}{\url{https://openbenchmark.github.io/BARS}}.


\subsection{Performance Evaluation with SOTA Models}

We evaluate the ReLoop2 module on existing models, including many state-of-the-art (SOTA) methods. The performances are shown in Table~\ref{tab:main_table}. Through the analysis of experiment results, we get some conclusions as follows: deep-learning-based methods get higher accuracy than the low-rank-based methods, thus revealing the powerful feature interaction ability of neural networks. In addition, xDeepFM obtains the second-best results on the MicroVideo dataset, indicating that a well-designed structure could fully use
the advantages of the factorization machine component. What's more, DIEN method obtains the second-best results on AmazonElectronics and KuaiVideo datasets, which benefits from the evolution of user interests and exploitation of the sequential features. In addition, we can see that our BASE+ReLoop2 outperforms all the other baseline methods since the error memory module is applied to the baseline method to augment the base encoder, and the error compensation helps to adapt to data distribution shift rapidly. Specifically, we choose DIEN as the base model for AmazonElectronics and KuaiVideo, and DCNv2 for MicroVideo because of their relatively better performance.
It is worth noting that our ReLoop2 framework is model agnostic to all the existing models, which is shown in Table~\ref{tab:agnostic}. 
After applying ReLoop2 to the five state-of-the-
art models respectively, we can obtain the new SOTA. 

\begin{figure}[t]
    \centering
    \includegraphics[scale=0.48]{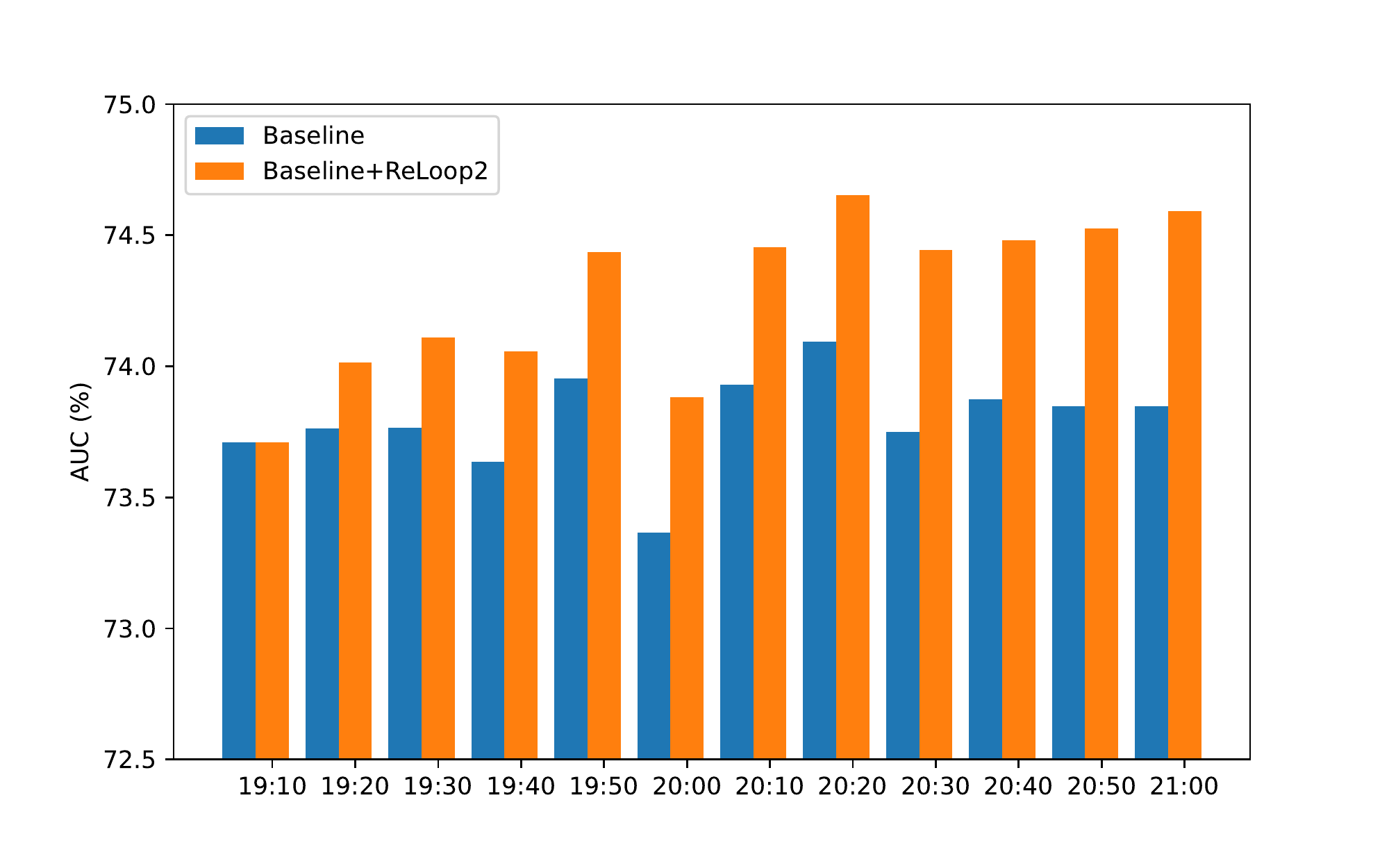}
    \caption{Evaluation of ReLoop2 on our production dataset.}
    \label{fig:product}
\end{figure}

\textbf{Evaluation on production dataset.} The comparison of our model with the baseline on the product dataset is shown in Figure~\ref{fig:product}. The baseline is an incremental learning method, which serves as the base encoder, so the results of the first part of the test set are exactly the same. From the second part, we utilize the previous parts as fast access error memory to learn the error compensation rapidly, and the performance demonstrates the efficiency of our ReLoop2 apporach.

\subsection{Evaluation between ReLoop2 and Incremental Training}

\begin{figure}[t]
    \centering
     \includegraphics[scale=0.26]{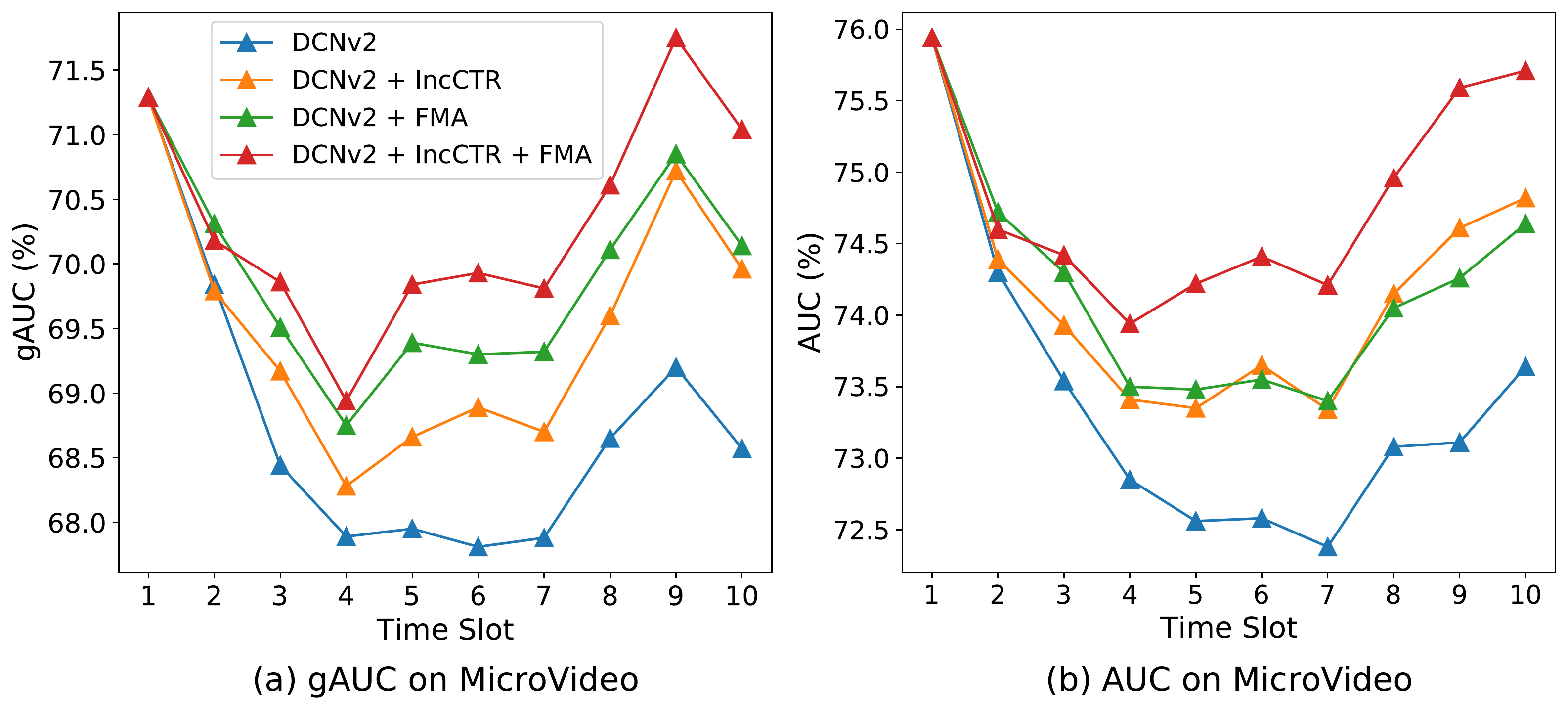}
    \caption{Comparison between ReLoop2 and incremental training on MicroVideo.}
    \label{fig:increment}
\end{figure}

As mentioned earlier, incremental model training has been a common choice in real-world production systems, so in Figure~\ref{fig:increment}, we compare our fast model adaptation with incremental training based on DCNv2 backbone on MicroVideo.
The horizontal axis of Figure~\ref{fig:increment} is time slot, as we split the test dataset of MicroVideo into
ten time slots in chronological order evenly to simulate online advertising task. Four methods are compared in Figure~\ref{fig:increment}.
\begin{itemize}[leftmargin=*]
\item \textbf{DCNv2} is the baseline model in this experiment.
\item \textbf{DCNv2+IncCTR} applies the incremental training method, IncCTR~\cite{IncCTR}, on top of DCNv2. Specifically, after model training on the training data and model evaluation on the first part of the test data, we continually train the model using the first part of the test data and then evaluate it on the second part. The process goes on like this on ten test parts. Note that we only pass the test data once for IncCTR as suggested in the paper.

\item \textbf{DCNv2+ReLoop2} applies fast model adaptation (FMA) to DCNv2.
\item \textbf{DCNv2+IncCTR+ReLoop2} applies both incremental training and fast model adaptation (FMA) to DCNv2. Note that our ReLoop2 framework is orthogonal to the incremental training technique since ReLoop2 do not need extra training.
\end{itemize}
In Figure~\ref{fig:increment}, ReLoop2 outperforms IncCTR most of the time on both gAUC and AUC, except for the last two time slots, where AUC of IncCTR exceeds that of ReLoop2.
It is understandable since, with the passage of time, new data distribution changes dramatically from the initial data distribution, and as a result, the accuracy of the original base model's prediction for new data decreases. 
As ReLoop2 relies on base model prediction and error memory with no training process, it is likely that the AUC of IncCTR exceeds that of ReLoop2 when the time slot increases. From another point of view, additional
training, like incremental learning, is necessary since it can make the model have better control over new data by updating the model parameters.
By combining IncCTR and ReLoop2, DCNv2 achieves the best performance in Figure~\ref{fig:increment}, demonstrating the efficiency of our fast model adaptation module.

\begin{figure}[!htbp]
    \centering
     \includegraphics[scale=0.26]{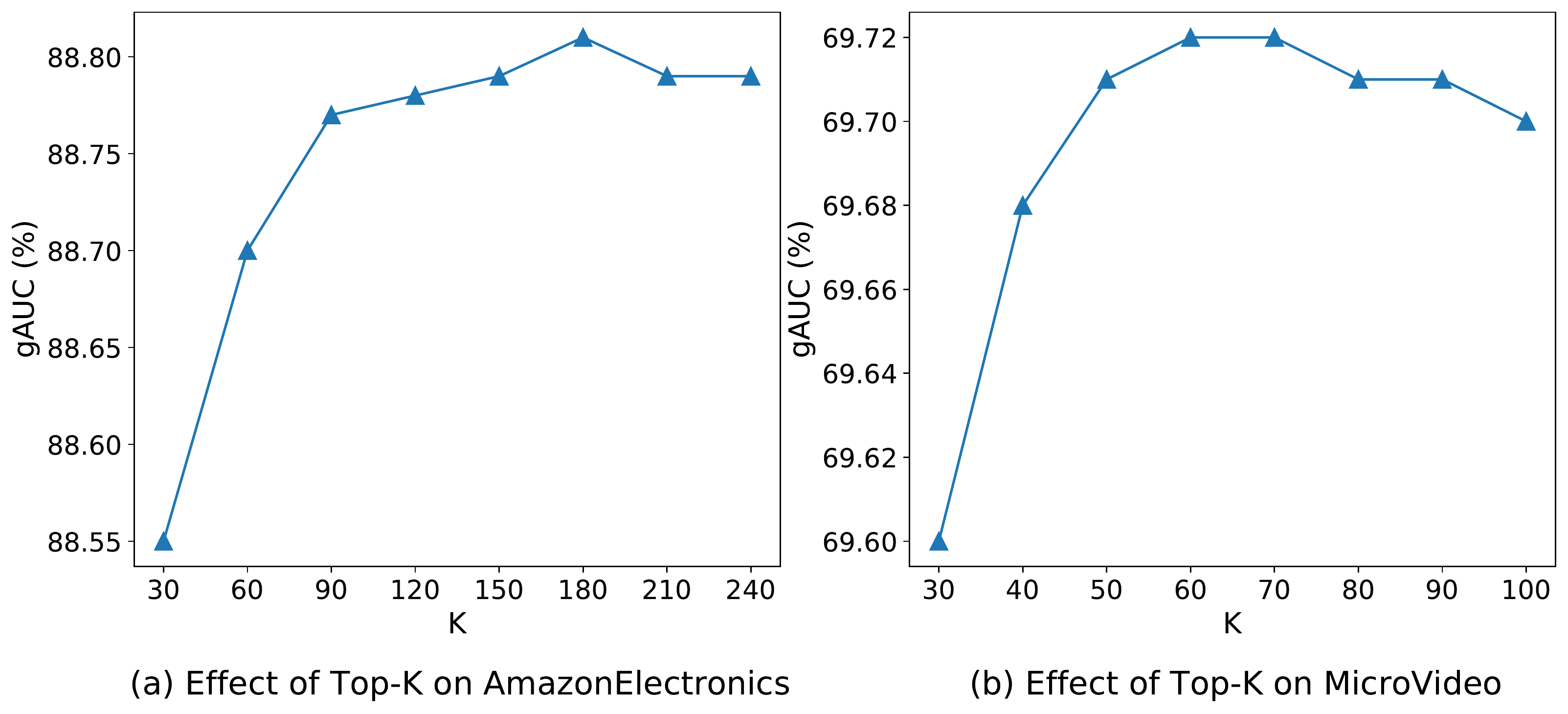}
    \caption{Effect of $K$ on AmazonElectronics and MicroVideo.}
    \label{fig:topk}
\end{figure}

\begin{figure}[!htbp]
    \centering
     \includegraphics[scale=0.26]{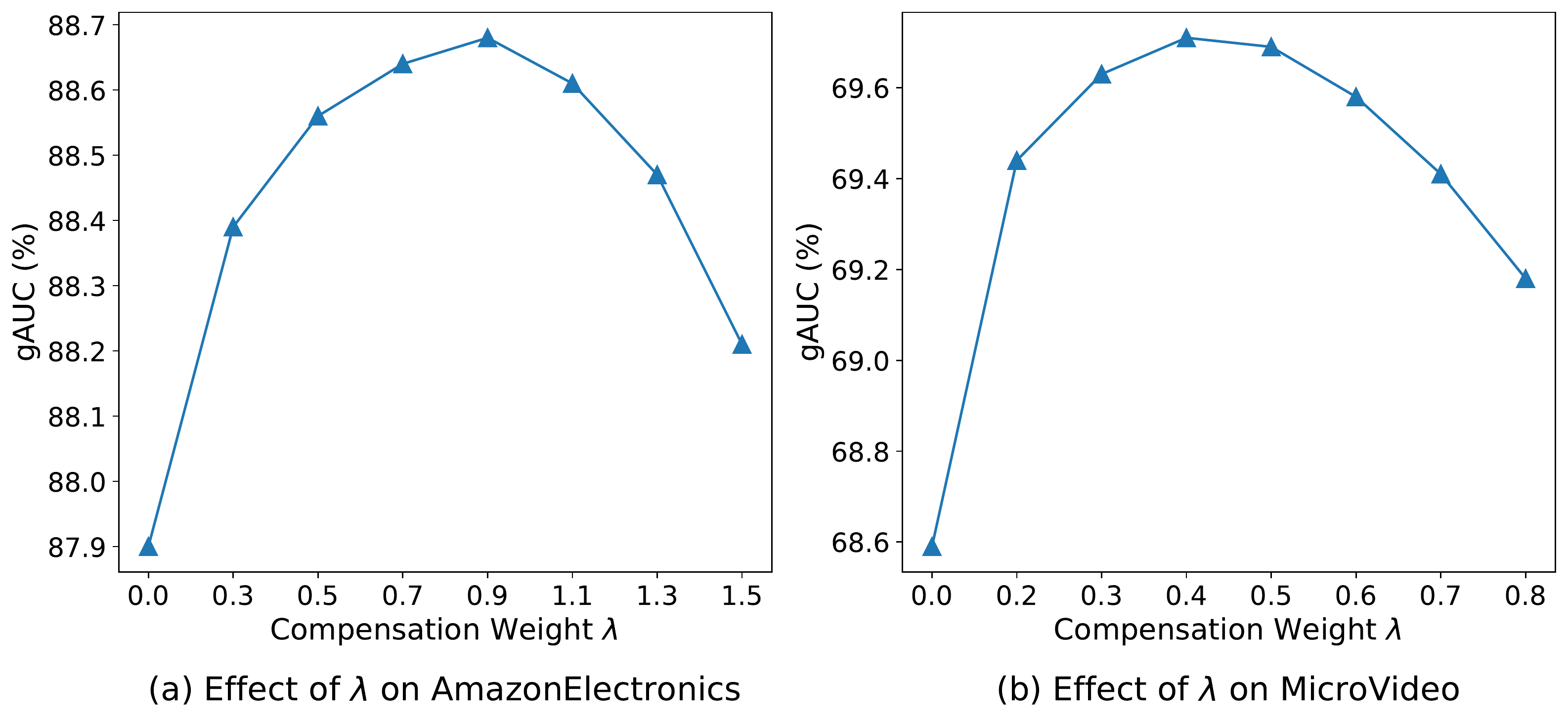}
    \caption{Effect of compensation weight $\lambda$ on AmazonElectronics and MicroVideo.}
    \label{fig:compensation}
\vspace{-1ex}
\end{figure}

\subsection{Ablation Studies}
\subsubsection{Effect of $K$ for  Memory Reading}

We investigate the effect of K in Figure~\ref{fig:topk}.
When K is small, error compensation is mainly determined by a small number of neighbors, which can not stand for the average error in the error memory, leading to a higher but not the best gAUC.
When K is too large, the final output will be influenced by those neighbor samples that are not so similar to itself, resulting in a slight decrease in gAUC, but it is generally stable.
The best results are obtained when we choose an appropriate K. Through our experiment, we find that K=180 and K=70 achieve the best performance of DCNv2 on AmazonElectronics and MicroVideo, respectively.

\subsubsection{Effect of Compensation Weight $\lambda$}

We investigate the effect of compensation weights $\lambda$ in Figure~\ref{fig:compensation}.
When $\lambda$ is small, the final
output is mainly determined by the base model output, thus a bit higher but relatively close to the baseline. Specifically, when $\lambda=0$, the final output is the same as that of the base model, serving as the baseline. 
When $\lambda$ is too large, error compensation contributes more to the final output. gAUC of the final output decreases because of the lower percentage of base model, whose accuracy is supported by a large amount of training data.
We empirically find $\lambda=0.9$ and $\lambda=0.4$ achieve the best performance of DCNv2 on AmazonElectronics and MicroVideo, respectively.

\section{Related Work}

\textbf{CTR Prediction}. CTR prediction plays a key role in online advertising, recommender system, and information retrieval. Even a small improvement in CTR prediction can have a significant impact, benefiting both users and platforms. As a result, extensive research efforts have been dedicated to this field, both in academia and industry. In general, the goal of CTR prediction is to generate probability scores that represent user preferences for item candidates in specific contexts. Recently, a plethora of CTR prediction approaches have been proposed, ranging from traditional logistic regression (LR) models~\cite{FTRL}, factorization machines (FM) models~\cite{FM,FFM}, to various deep neural network (DNN) models. Many of these models focus on designing feature interaction operators to capture complex relationships among features, such as product operators~\cite{PNN, NFM, DBLP:conf/www/SunPZF21}, bilinear interaction and aggregation~\cite{fibinet,finalmlp}, factorized interaction layers~\cite{FINAL}, convolutional operators~\cite{ccpm, CNN-FeatureGen, FiGNN}, and attention operators~\cite{BST, autoint}. Additionally, user behavior sequences play a crucial role in modeling user interests. Different models have been employed for behavior sequence modeling, including attention-based models~\cite{DIN,DIEN}, memory network-based models~\cite{HPMN,UIC}, and retrieval-based models~\cite{UBR,SIM}. Notably, the BARS benchmark~\cite{FuxiCTR,BARS} provides a comprehensive review and benchmarking results of existing CTR prediction models. However, all of these models focus on modeling sequential patterns during training and assume fixed parameters during testing, making them incapable of handling distribution shifts. 

\textbf{Incremental Learning}. 
Incremental learning is a general framework that aims to continuously update model parameters to acquire new knowledge from new data while preserving the model's ability to generalize on old data~\cite{IncLearning}. In the context of recommender systems, incremental training has been widely adopted to cope with the data distribution shift and minimize the generalization gap between training and testing. Typically, model parameters from the previous version are reused as an initialization for the next round of training~\cite{IncrementalCTR}. To alleviate the catastrophic forgetting problem, Wang et al.~\cite{IncCTR} proposed the IncCTR method, which uses knowledge distillation to strike a balance between retaining the previous pattern and learning from the new data distribution. In our earlier work, we introduced the ReLoop framework\cite{ReLoop}, which establishes a self-correcting learning loop during the model training phase. However, ReLoop2 focuses on test-time adaptation instead. Integrating both techniques is an interesting direction and we leave it for future research. Other studies~\cite{MetaInc1, MetaInc2,IncrementalRS} apply meta-learning techniques to incremental training of recommendation models, aiming to facilitate knowledge transfer from old parameters to new parameters. A recent study~\cite{ConceptDrift} proposes an adaptive incremental learning algorithm for mixture-of-experts (MoE) models to adapt to concept drift. Instead, our work is orthogonal to incremental training and focuses on enabling fast model adaptation through error compensation using a non-parametric memory approach. Furthermore, in contrast to the majority of continual learning studies~\cite{CLSurvey}, ReLoop2 employs a refreshed error memory for model adaptation, deviating from the conventional practice of utilizing a memory buffer for experience replay to prevent catastrophic forgetting.

\textbf{Retrieval Augmentation}. 
Our work also draws some inspiration from recent research on retrieval augmented machine learning techniques~\cite{RAML}. Retrieval augmentation focuses on retrieving similar key-value pairs from the external memory to enhance model generalizability, particularly for rare events or long-tail classes~\cite{RA_Mem}. This approach has been successfully applied in various domains, including neural machine translation~\cite{KNN_NMT, Fast_KNN_NMT, Faster_KNN_NMT}, visual recognition~\cite{RA_Classification,RA_Classification2}, question answering~\cite{RAG}, retrieval-augmented pre-training~\cite{REALM,REVEAL} and text-to-image generation~\cite{RA_Diffusion}. However, unlike these studies that retrieve data for model training, our work leverages refreshed online data for retrieval-augmented model adaptation. Additionally, we present a time- and memory-efficient design for top-k retrieval in large-scale online recommendation scenarios.

\section{CONCLUSION}
In this paper, we make a pioneering effort towards fast adaptation of CTR prediction models for online recommendation. To address the challenge of distribution shifts in streaming data, we introduce a slow-fast learning paradigm inspired by the complementary learning systems observed in human brains. In line with this paradigm, we propose ReLoop2, a self-correcting learning loop that facilitates fast model adaptation in online recommender systems through responsive error compensation. Central to ReLoop2 is a non-parametric error memory module that is designed to be time- and space-efficient and undergoes continual refreshing with newly observed data samples during model serving. Through comprehensive experiments conducted on open benchmark datasets and our production dataset, we demonstrate the effectiveness of ReLoop2 in enhancing model adaptiveness under distribution shifts. 


\section*{Acknowledgments}
We gratefully acknowledge the support of MindSpore\footnote{\url{https://www.mindspore.cn}}, which is a new deep learning computing framework used for this research.

\balance
\bibliographystyle{ACM-Reference-Format}
\bibliography{ReLoop2}
\appendix
\end{document}